\newcommand{\be}{\begin{equation}}\newcommand{\ee}{\end{equation}}
\newcommand{\bea}{\begin{eqnarray}}\newcommand{\eea}{\end{eqnarray}}
\newcommand{\nn}{\nonumber}\newcommand{\p}[1]{(\ref{#1})}
 \newcommand{\lb}[1]{\label{#1}}
\newcommand\q{\quad}\newcommand\qq{\qquad}
\newcommand\tp{\theta^+}
\newcommand\btp{\bar{\theta}^+}

\newcommand\tpb{\theta^{\beta+}}
\newcommand\tpa{\theta^{\alpha+}}

\newcommand\btpa{\bar{\theta}^{\dot{\alpha}+}}
\newcommand\btpb{\bar{\theta}^{\dot{\beta}+}}

\newcommand\tm{\theta^-}
\newcommand\btm{\bar{\theta}^-}
\newcommand\tma{\theta^{\alpha-}}
\newcommand\tmb{\theta^{\beta-}}

\newcommand\btma{\bar{\theta}^{\dot{\alpha}-}}
\newcommand\btmb{\bar{\theta}^{\dot{\beta}-}}

\newcommand\Tpb{\Theta^{\beta+}}
\newcommand\Tpa{\Theta^{\alpha+}}
\newcommand\hT{\hat\Theta}
\newcommand\hTp{\hat\Theta^+}
\newcommand\hTpa{\hat{\Theta}^{\dot{\alpha}+}}
\newcommand\hTpb{\hat{\Theta}^{\dot{\beta}+}}

\newcommand\hTm{\hat{\Theta}^-}
\newcommand\Tma{\Theta^{\alpha-}}

\newcommand\hTma{\hat{\Theta}^{\dot{\alpha}-}}

\newcommand\pada{\partial_{\alpha\dot{\alpha}}}
\newcommand\padb{\partial_{\alpha\dot{\beta}}}

\newcommand\pbdb{\partial_{\beta\dot{\beta}}}

\newcommand\ada{{\alpha\dot{\alpha}}}
\newcommand\adb{{\alpha\dot{\beta}}}
\newcommand\bdb{{\beta\dot{\beta}}}
\newcommand\bda{{\beta\dot{\alpha}}}
\newcommand\ab{{\alpha\beta}}

\newcommand\bg{{\beta\gamma}}

\newcommand\ar{{\alpha\rho}}

\newcommand\br{{\beta\rho}}

\newcommand\da{{\dot{\alpha}}}
\newcommand\db{{\dot{\beta}}}

\newcommand\cA{{\cal A}}
\newcommand\cB{{\cal B}}
\newcommand\cV{{\cal V}}
\newcommand\cD{{\cal D}}

\newcommand\hcD{\hat{\cal D}}
\newcommand\hcA{\hat{\cal A}}
\newcommand\hcW{\hat{\cal W}}
\newcommand\cW{{\cal W}}

\newcommand\s{\scriptscriptstyle}
\newcommand\A{{\s A}}
\newcommand\B{{\s B}}
\newcommand\C{{\s C}}
\newcommand\R{{\s R}}
\newcommand\M{{\s M}}

\newcommand{\pp}{{\s ++}}
\newcommand{\m}{{\s --}}

\newcommand{\Dp}{D^{\pp}}
\newcommand{\Dm}{D^{\m}}
\newcommand{\Vp}{V^{\pp}}
\newcommand{\Vm}{V^{\m}}

\newcommand{\dpp}{\partial^{\pp}}
\newcommand{\dm}{\partial^{\m}}
\newcommand{\DP}{D^+}

\newcommand{\DPa}{D^+_\alpha}

\newcommand{\DMa}{D^-_\alpha}

\newcommand{\pPa}{\partial^+_\alpha}

\newcommand{\pMa}{\partial^-_\alpha}

\newcommand{\bDP}{\bar{D}^+}
\newcommand{\bDPa}{\bar{D}^+_{\dot{\alpha}}}
\newcommand{\bDPb}{\bar{D}^+_{\dot{\beta}}}

\newcommand{\bDMa}{\bar{D}^-_{\dot{\alpha}}}

\newcommand{\bpPa}{\bar{\partial}^+_{\dot{\alpha}}}

\newcommand{\bpMa}{\bar{\partial}^-_{\dot{\alpha}}}

\newcommand{\cDpa}{\cD^+_\alpha}

\newcommand{\cDma}{\cD^-_\alpha}

\newcommand{\hcDpa}{\hcD^+_\da}
\newcommand{\hcDpb}{\hcD^+_\db}

\documentclass[12pt]{article}
\def\theequation{\arabic{section}.\arabic{equation}}

\begin{document}
\renewcommand{\thefootnote}{\fnsymbol{footnote}}
\begin{flushright}
{ hep-th/0107012 }
\end{flushright}
\vspace{2cm}

\begin{center}
{\large\bf   GEOMETRY OF SOLUTIONS OF
N=2 SYM-THEORY IN HARMONIC SUPERSPACE}
\vspace{1cm} \\

{\bf B.M. Zupnik}\\
\vspace{0.5cm}

{\it Joint Institute for Nuclear Research, Bogoliubov Laboratory
of Theoretical Physics, Dubna, Moscow Region,
 141980, Russia. E-mail: zupnik@thsun1.jinr.ru }

\end{center}

\begin{abstract}
In harmonic superspace, the classical equations of motion of $D=4, N=2$
supersymmetric Yang-Mills theory for Minkowski and Euclidean spaces are
analyzed. We study dual superfield representations of equations and
subsidiary conditions corresponding to classical SYM-solutions with
different symmetries. In particular, alternative superfield constructions
of self-dual  and static  solutions are described in the framework
of the harmonic approach.
\end{abstract}

\setcounter{equation}0
\section{Introduction}
The off-shell superfield constraints of the $N=2 $ super-Yang-Mills theory
in the Minkowski $D=(3,1)$ space has been solved in the framework of
harmonic superspace (HSS) using the auxiliary  coordinates of the coset
space $S_2\sim SU_\A(2)/U_\A(1)$ where $SU_\A(2)$  is the corresponding
automorphism group \cite{GIK1}-\cite{Zu2}. An analogous harmonic formalism
can be applied to reformulate the off-shell $N=2$ supersymmetric models in
the Euclidean $D=4$ space. We shall use the notation with the subscripts
$HSS_M$ or $HSS_E$ for harmonic-superspace structures over Minkowski or
Euclidean  spaces, respectively. Auxiliary harmonic coordinates are used
in  covariant conditions of the Grassmann (G-) analyticity. The
unconstrained superfields of the $N=2$ theories live in the G-analytic
superspaces with the reduced odd dimension $4$.

Harmonic variables connected with coset $SU_\R(2)/U_\R(1)$ of the
subgroup of the Euclidean rotation group have been used to study self-dual
solutions of the YM- and  SYM-theories \cite{GIOS2,OO,DO}. These harmonic
superspaces have the reduced values of even and odd dimensions, and the
corresponding analytic superfields parametrize moduli spaces of self-dual
solutions.

The  superfield equations of motion in the most general harmonic
formalism can  be effectively used for the analysis of the geomery of
classical $N=2$ SYM-solutions. In particular, one can consider  dual
transformations of the $N=2$ SYM-superfield variables in the harmonic
superspace which allow us to formulate unusual representations of the
equations of motion and simple gauge conditions \cite{Zu3,Zu4}. It is
important to stress that the dual change of variables transforms the part
of SYM-constraints and equations in the harmonic formalism  to linear
restrictions on HSS-connection  $\Vm$, then the zero-curvature relation
between the harmonic connections can be interpreted as a dynamical
equation. Dual representations of the HSS-superfield equations allow us to
describe symmetry properties of different classes of partial solutions.

It is interesting to compare the alternative HSS-descriptions of $4D$
self-dual SYM-solutions or solutions of $D<4$ SYM-equations. We analyze
the  $(t=0)$ reduction of the $HSS_M$-formalism and formulate the static
3-dimensional superfield BPS-conditions. It is shown that the
supersymmetry of the static self-dual SYM-equations is equivalent to the
corresponding  3-dimensional subgroup of the Euclidean supersymmetry
with 8 supercharges.

The $HSS_M$-equations of the $N=2$ SYM-theory are considered in Sect. 2.
We discuss the off-shell formalism with independent harmonic connections
and dually-equivalent constructions of the superfield equations of motion.
One analyzes the conditions of the G-analyticity for connections in this
formalism. The superfield equations of motion can be transformed to
relations between the G-analytic functions arising in decompositions of
the non-analytic connection $\Vm$.

Sect. 3 is devoted to the discussion of the static $3D$ reduction of the
$N=2$ superfield SYM-equations. The time component of the $4D$ superfield
connection $A_t$ becomes the new $3D$-scalar superfield in this limit. We
consider the $3D$-superfield BPS-type relation between $A_t$ and the
superfield strengths $W$ and $\bar{W}$ which is equivalent to the 2nd
order differential constraint for  connection $\Vm$. It is shown that this
2nd order constraint generates the standard 4th order constraint for the
same connection  and all solutions of the superfield BPS-equation satisfy
the SYM-equation.

In Sect. 4 we study the $4D$ Euclidean version of the harmonic $HSS_E$
formalism. The chiral superfield self-duality condition in this approach
can also be interpreted as the 2nd order constraint on $\Vm$. The
alternative bridge representation and the nilpotent gauge condition
for the bridge superfield are used  to analyze the $HSS_E$ self-dual
solutions.

The comparison of these harmonic constructions with the analogous
self-dual $N=2$ solutions in the alternative $SU_\R(2)/U_\R(1)$ formalism
is discussed in Sect. 5. This version of the harmonic approach transforms
the super-self-duality condition to the specific Grassmann-bosonic
analyticity of the on-shell harmonic connection. We consider the
identification of the harmonic variables for different $SU(2)$ subgroups
and the simple Ansatz for the self-dual harmonic connection in the gauge
group $SU(2)$ which yield the explicit construction of solutions in
quadratures \cite{Zu}.

The static self-dual solutions are discussed in Sect. 6. We discuss the
relations between the conjugated spinor coordinates in the static limit of
the $4D$  superspace $HSS_M $and the corresponding pseudoreal spinor
coordinates of the $3D$ reduction of the Euclidean  superspace $HSS_E$.

The basic formulae of the harmonic-superspace approach are reviewed
in Appendix. We describe conjugation rules for different versions of
harmonic superspaces which are very important in the  SYM-theory.

\setcounter{equation}0
\section{\lb{B}Harmonic-superspace representations  of N=2 SYM-equations}

The harmonic $SU_\A(2)/U_\A(1)$ transform has been used first to solve
the off-shell  constraints for  superfield connections $A^k_a(z)$ of
the $N=2, D=(3,1)$ SYM-theory \cite{GIK1,GIOS}
\be
u^+_k(D^k_a+A^k_a)\equiv\nabla^+_a= e^{-v}D^+_a e^v
\ee
where $a=(\alpha,\da)$ are the $SL(2,C)$ indices, $k$ is the 2-component
index of automorphism group $SU(2)_\A$, $u^+_k$ are the auxiliary
harmonic variables (see Appendix) and $v(z,u)$ is the {\it bridge} matrix.
This transform connects formally different off-shell representations of
the gauge group in the central basis (CB) and the analytic basis (AB).

The basic off-shell harmonic superfield of the  SYM-theory is the
connection $\Vp=e^v\Dp e^{-v}$ for the harmonic covariant derivative of
the analytic basis
\be
\nabla^\pp=\Dp+\Vp~,\q \delta \Vp=-\Dp\lambda-[\Vp,\lambda]~.\lb{B1}
\ee
The prepotential $\Vp$ and the AB-gauge parameters satisfy
the G-analyticity conditions
\be
(\DPa, \bDPa)(\Vp, \lambda)=0~.\lb{B2}
\ee

The G-analytic superfields are described by the unconstrained functions
of the analytic coordinates $\zeta$ \p{F5} and harmonics $u^\pm_i$.
The second harmonic connection $\Vm$ satisfies the harmonic zero-curvature
 equation
\be
\Dp \Vm-\Dm\Vp+[\Vp,\Vm]=0~,\lb{B3}
\ee
which can be solved explicitly in terms of $\Vp$-serias.

In the analytic basis, the spinor and vector connections $A_\M$ as well as
the superfield strengths $W$ and $\bar{W}$ can be written directly via
spinor derivatives of connection $\Vm$ \cite{Zu2}
\bea
&&A^+_\alpha=\bar{A}^+_\da=0~,\q A^-_\alpha=-\DPa\Vm~,\q
\bar{A}^-_\da=-\bDPa\Vm~,\\
&&A_\adb\equiv i\DPa\bar{A}^-_\db\equiv-i\bDPb A^-_\alpha=-i\DPa\bDPb\Vm~,
\lb{B5b}\\
&& \bar{W}={1\over2}D^{\alpha+}A^-_\alpha=-(\DP)^2\Vm~,\q
W=-{1\over2}\bDP_\da\bar{A}^{\da-}=(\bDP)^2\Vm~.\lb{B5}
\eea

The nonlinear harmonic-superfield equation of motion for the prepotential
$\Vp$ can be reformulated as a linear differential condition on the
composed connection $\Vm(\Vp)$
\be
(\DP)^2(\bDP)^2\Vm=0~.\lb{B6}
\ee
 Eq.\p{B3} is treated as  a kinematic solvable equation in the standard
SYM-formalism with basic superfield $\Vp$. The alternative $HSS_M$
formalism with the independent off-shell superfield $\Vm$ has been
considered in Refs.\cite{Zu3}. This approach is analogous to the
first-order formalism for the ordinary Yang-Mills theory, since the
independent superfield $\Vm$ contains off-shell an infinite number of
auxiliary fields including the fields of dimension $-2:~ F, F_{mn}$ and
$F_m$. (Note that the changes of basic HSS variables $\Vp\rightarrow \Vm$
or $\Vp\rightarrow v$ can be interpreted as dual transformations in the
SYM-theory. We hope that the rich  structure of the dual
transformations in HSS will be useful for the explicit constructions
of different classical solutions.)

The G-analyticity of $\Vp$ is equivalent to the  nonlinear
analyticity  equation for $\Vm$
\be
[\nabla^-_a,\nabla^\m]=D^-_a\Vm+\Dm D^+_a\Vm+[\Vm,D^+_a\Vm]=0~.\lb{B18}
\ee

The linear constraint \p{B6} for the independent superfield $\Vm$ can be
readily solved  \cite{Zu3}
\be
\Vm=\DPa B^{\alpha(-3)}-\bDPa\bar{B}^{\da(-3)}\lb{B31b}.
\ee

In the picture with independent SYM superfield variables
$B^{\alpha(-3)}$ and $\Vp$, the basic dynamical equation is
\be
\nabla^\pp(\DPa B^{\alpha(-3)}-\bDPa\bar{B}^{\da(-3)})=\Dm\Vp\lb{B32b}~.
\lb{Beq}
\ee

Acting by  operators $\DPa ,\bDPa$  on both sides of this
equation one can obtain the following relations:
\bea
&&\nabla^\pp(\bDP)^2\DPa B^{\alpha(-3)}=0~,
 \lb{B33b}\\
&&\nabla^\pp
[(\bDP)^2\DPa\bar{B}_\da^{(-3)}-(\DP)^2\bDPa B_\alpha^{(-3)}]=i\pada\Vp~.
\eea

Decomposition of Eq.\p{B32b} in terms of all spinor coordinates
gives  the component SYM-equations; however, it is also useful  to analyze
the partial decomposition of this superfield equation in terms of $\tma$
and $\btma$. Let us define the G-analytic components of  superfield $\Vm$
\bea
&\bar\phi=-(\DP)^2\Vm |~,\q \phi=(\bDP)^2\Vm |~,\q B_\adb=\bDPb\DPa\Vm |
~,&\lb{B7}\\
&\lambda_\alpha^+=-\DPa(\bDP)^2\Vm |~,\q\bar\lambda_\da^+=
-\bDPa(\DP)^2\Vm |~,\q F^\pp=(\DP)^2(\bDP)^2\Vm |~,&\nn
\eea
where the symbol $|$ means $\theta^-=\bar\theta^-=0$.

The full supersymmetry transformation $\delta_\epsilon \Vm=0$ yields
the corresponding transformations of the analytic components
\bea
&&\delta_\epsilon \phi=u^-_k\epsilon^{\alpha k}\lambda_\alpha^+~,\q
\delta_\epsilon B_\adb=u^-_k\epsilon_\alpha^k\bar\lambda_\db^+
+u^-_k\bar\epsilon_\db^k\lambda_\alpha^+~,\lb{B9}\\
&&\delta_\epsilon\lambda_\alpha=-u^-_k\epsilon_\alpha^k F^\pp~,\q
\delta_\epsilon F^\pp=0~.
\eea

Using the gauge transformation $\delta \Vm=-\Dm\lambda-[\Vm,\lambda]$,
one can obtain the nilpotent gauge condition for the harmonic
connection
\bea
&&\cV^\m=(\tm)^2\bar\phi-(\btm)^2\phi+\tma\btmb B_\adb+
(\tm)^2\btma\bar\lambda^+_\da+(\btm)^2\tma\lambda^+_\alpha\nn\\
&& +
(\tm)^2(\btm)^2F^\pp~.\lb{B9b}
\eea

The geometric superfields have the following form in this gauge:
\bea
&&W=\phi-\tma\lambda^+_\alpha-(\tm)^2F^\pp~,\lb{gauge1}\\
&&\bar{W}=\bar\phi+\btma\bar\lambda^+_\da+(\btm)^2F^\pp~,\lb{gauge2}\\
&&A_\adb=-iB_\adb+i\tm_\alpha\bar\lambda^+_\db+i\btm_\db\lambda^+_\alpha
-i\tm_\alpha\btm_\db F^\pp\lb{gauge3}~.
\eea

The constraint  $F^\pp=0$ gives the following  analytic equations
of motion equivalent to Eq.\p{B32b}
\bea
&&\nabla^\pp\phi=\tpa\lambda^+_\alpha~,\q(\nabla^\pp)^2\phi=0~,\lb{B17b}
\\
&&\nabla^\pp B_\adb=-i\padb\cV^\pp+\theta^+_\alpha\bar\lambda^+_\db+
\bar\theta^+_\db\lambda^+_\alpha~,\lb{B17c}\\
&&\nabla^\pp\lambda^+_\alpha=0
~.\lb{B30b}
\eea

The component solutions for the on-shell $N=2$ analytic superfields are
\bea
&&\phi=\varphi(x)+\tpa u^-_k\psi^k_\alpha(x)~,\\
&&B_\adb=a_\adb(x)+\theta^+_\alpha u^-_k\bar\psi_\db^k(x)+
\bar\theta^+_\db u^-_k\psi^k_\alpha(x)~,\\
&&\cV^\pp=(\tp)^2\bar\varphi(x)-(\btp)^2\varphi(x)+\tpa\btpb a_\adb(x)-
\tpa(\btp)^2 u^-_k\psi^k_\alpha(x)\nn\\
&&-\btpa(\tp)^2 u^-_k\bar\psi^k_\da(x)~.\lb{D1}
\eea

The  $N=2$ equations of motion fot these components follow from
G-analytic superfield equations. Of course, the main purpose of the
harmonic approach is to find nonstandard superfield methods of solving the
SYM-equations which cannot be completely reduced to the analysis of
component equations. We hope that a comparison of dual superfield
representations of  SYM-solutions will be useful for the search of unusual
symmetry properties which cannot be seen directly in the component
representation.

\setcounter{equation}0
\section{\lb{C}Static superfield equations}

Three-dimensional monopole and dyon solutions play an important role in
modern non-perturbative methods of the quantum $N=2$ SYM-theory
\cite{SW,Ga}. The harmonic analysis of the monopole Yang-Mills solutions
has been considered in Ref.\cite{OO}. We shall analyze the alternative
harmonic-superfield constructions of the static $N=2$ SYM-solutions and
their relations with the $4D$ self-dual solutions.

Let us consider the nonrelativistic representation of the coordinates in
the $D=(3,1),N=2$ harmonic superspace based on the static group $SO(3)$
\bea
&&x^\adb_\A\rightarrow iy^\alpha_\beta+\delta^\alpha_\beta
 t_\A~,\q y^\alpha_\alpha=0~,\q y^\ab=\varepsilon^\ar y^\beta_\rho
~,\lb{C1}\\
&&\padb~\rightarrow~ -i\partial^\alpha_\beta+\delta_\alpha^\beta
\partial_t~,\q\partial_\gamma^\sigma y^\alpha_\beta=2\delta^\sigma_\beta
\delta^\alpha_\gamma-\delta^\alpha_\beta\delta^\sigma_\gamma~,\\
&&\bar\theta^{\da\pm}~\rightarrow~\bar\theta^\pm_\alpha~,\q
\bar\theta^{\alpha\pm}=\varepsilon^\ab\bar\theta^\pm_\beta~,\\
&&\bar\partial^\pm_\da~\rightarrow~\bar\partial^{\alpha\pm}~,\q
\bar\partial^{\alpha\pm}\bar\theta_\beta^\mp=\delta^\alpha_\beta~.
\eea

We shall use the following conjugation rules:
\bea
&&(y^\alpha_\beta)^\dagger=-y_\alpha^\beta~,\q
(y^\ab)^\dagger=y_\ab~,\q(t_\A)^\dagger=t_\A~,\\
&&(\theta^{\alpha\pm})^\dagger= \bar\theta_\alpha^\pm~,\q
(\theta_\alpha^\pm)^\dagger= -\bar\theta^{\alpha\pm}~,\q
(\varepsilon^\ab)^\dagger=-\varepsilon_\ab~,\\
&&(\bar\theta^{\alpha\pm})^\dagger= \theta_\alpha^\pm~,\q
(\bar\theta_\alpha^\pm)^\dagger= -\theta^{\alpha\pm}~,\lb{C2}\\
&&(D_\alpha^\pm)^\dagger= \bar{D}^{\alpha\pm}~,\q
(\bar{D}^{\alpha\pm})^\dagger=-D_\alpha^\pm~,\q
(\bar{D}_\alpha^\pm)^\dagger=D^{\alpha\pm}~.
\eea
Note that changes of the position of dotted $SL(2,C)$ indices after the
static reduction is connected with a convention of the conjugation of the
$SU(2)$-spinors which transforms upper indices to lower ones. The time
reduction $(t=0)$ transforms 4-dimensional harmonic superfields to the
3-dimensional Euclidean  superfields which are covariant with respect to
the reduced supersymmetry with 8 supercharges.

The corresponding representation of the  4-vector connection is
\be
A_\adb~\Rightarrow~A^\beta_\alpha+\delta^\beta_\alpha A_t~,\q
(A^\beta_\alpha)^\dagger=-A_\beta^\alpha~,\q(A_t)^\dagger=-A_t~,
\lb{C3b}
\ee
where the time component of the connection  becomes covariant with
respect to residual gauge static transformations
\bea
&&A^\beta_\alpha=i(\bar{D}^{\beta+}\DPa-{1\over2}\delta^\beta_\alpha
(\bar{D}^{+}\DP) )\Vm~,\\
&&A_t={i\over2}(\bar{D}^{+}\DP) \Vm~,
\q\delta_\lambda A_t=[\lambda, A_t]~.\lb{C4}
\eea

It seems useful to construct the manifestly supersymmetric
generalization of the Bogomolnyi equation for the $N=2$ gauge theory
in order to analyze the symmetry properties of the corresponding
monopole solutions. In addition to the $t=0$ reduction of the HSS-equation
\p{B3}, we propose to consider the following superfield BPS-type relation:
\be
S\equiv-2irA_t+p(W+\bar{W})=0~,\lb{C6}
\ee
where $r$ and $p$ are some real parameters. Note that this condition
breaks  the $R$-symmetry $W\rightarrow e^{i\rho}W$.

We shall interpret \p{C6} as a linear constraint on $\Vm$
\be
S= [r(\bDP \DP)+p(\bDP)^2-p(\DP)^2]\Vm =0~,
\ee
which gives also the spinor condition
\be
\left(-r(\DP)^2\bar{D}^+_\alpha+p(\bDP)^2\DPa\right)\Vm=0~.\lb{C7}
\ee

For the case $r\neq p$ , the last spinor relation is not self-adjoint,
so one can obtain the strong restrictions
\be
\DPa(\bDP)^2\Vm=\DPa W=0~\Rightarrow~\nabla^\m\DPa W=\nabla^-_\alpha W=0
~.\lb{gen}
\ee
It is evident that the corresponding  covariantly constant solution
$W=const$ preserves all 8 supercharges \cite{IKZ}.  The case $r=p$
corresponds to the self-adjoint  relation for 3D spinors
\be
\lambda^\alpha_k=\bar\lambda^\alpha_k=\varepsilon^\ab\varepsilon_{kl}
(\lambda^\beta_l)^\dagger~.
\ee
The corresponding self-dual static solutions  will be discussed in
Sect. 6.

\setcounter{equation}0
\section{\lb{D}Euclidean $N=2$ SYM-equations}

We shall analyze the superfield constraints of the $N=2$ SYM-theory in the
Euclidean superspace
\bea
&&\{\nabla^k_\alpha,\nabla^l_\beta\}=\varepsilon_\ab\varepsilon^{kl}
\cW~,\lb{E1}\\
&&\{\hat\nabla^k_\da,\hat\nabla^l_\db\}=\varepsilon_{\da\db}
\varepsilon^{kl}\hcW~,\lb{E2}\\
&&\{\nabla^k_\alpha,\hat\nabla^l_\db\}=\varepsilon^{kl}\nabla_\adb~,
\lb{E3}
\eea
where  $\cW$ and $\hcW$ are independent real superfield strengths and
indices $\alpha, \beta; \da, \db$ and $k, l$ describe, respectively, the
2-spinor representations of the Euclidean group $SU_{\s L}(2)\times
SU_\R(2)$ and the automorphism group $SU_\A(2)$.

Superspace analogues of the Euclidean self-duality equations have been
considered in Refs.\cite{Se}-\cite{Si}. We shall analyze the $N=2$
self-dual solutions in the framework of the Euclidean version of the
harmonic superspace $HSS_E$ (see Appendix). Underline that conjugation
rules of the  $N=2$ superspace coordinates and derivatives are
essentially different for the cases of the Euclidean group $SO(4)$ and the
Lorentz group $SO(3,1)$.

The $SU_\A(2)/U_\A(1)$ harmonic-superspace description of the Euclidean
$N=2$ SYM-theory is  similar to the corresponding formalism in $HSS_M$,
in particular, one can  use the same notation $\Vp, \Vm$ for the Euclidean
harmonic connections. The Euclidean spinor and vector connections of the
analytic basic have the following form:
\bea
&&\cA^+_\alpha=\hcA^+_\da=0~,\q \cA^-_\alpha=-\cDpa\Vm~,\q
\hcA^-_\da=-\hcDpa\Vm~,\lb{E4}\\
&&\cA_\adb\equiv \cDpa\hcA^-_\db\equiv\cDpa\hcDpb\Vm~.\lb{E5}
\eea

The reality of the  Euclidean superfield strengths is evident in $HSS_E$
\be
\cW=(\cD^+)^2\Vm~,\q\hcW=(\hcD^+)^2\Vm~.\lb{E6}
\ee

The system of self-duality equations has the
following form in the harmonic-superspace:
\be
\cD^\pp\Vm+[\Vp,\Vm]=\cD^\m\Vp~,\q(\cD^+)^2\Vm=0~.\lb{E7}
\ee
The manifest solution of the 2nd constraint
\be
\Vm=\cDpa \cB^{\alpha(-3)}~,\lb{E8}
\ee
yields the simple relation between $\cB^{\alpha(-3)}$ and $\Vp$
\be
\cDpa\nabla^\pp\cB^{\alpha(-3)}=\cD^\m\Vp~.
\ee

The self-duality equation is equivalent to the relation
\be
\cDpa\cA^-_\beta=0~.\lb{E9}
\ee

Using the equation $[\nabla^\m,\nabla^-_\alpha]=0$ one can derive the
chirality relations
\be
\{\nabla^\pm_\alpha,\nabla^\pm_\beta\}=0~.\lb{E10}
\ee

Let us reformulate now the self-duality condition in the bridge
representation of the harmonic connection \cite{GIK1}
\be
\Vm\equiv e^v\cD^\m e^{-v}~.\lb{E10b}
\ee
The  condition \p{E9} for  $\Vm$ corresponds to the partial analyticity
condition for the self-dual bridge matrix
\be
\cD^+_\alpha v=0~\Rightarrow~-\cD^+_\alpha \left(e^v\cD^\m e^{-v}\right)
=e^v\cD^-_\alpha e^{-v}~.
\ee

Comparing representations \p{E8} and \p{E10b} one can obtain the relation
\be
\cB^{\alpha(-3)}={1\over2}\Theta^{\alpha-} e^v\cD^\m e^{-v}+
{1\over2}(\Theta^-)^2 e^v\cD^{\alpha-} e^{-v}~.
\ee

Harmonic projections of the central-basis connections
have the following form in this representation:
\bea
&&A_\alpha^\pm=0~,\q \hat{A}_\da^+(v)=e^{-v}\hcD_\da^+e^v~,\lb{vrep}\\
&&\hat{A}_\da^-=\cD^\m\hat{A}_\da^+(v)~,\q A_\ada(v)=\cD^-_\alpha
\hat{A}_\da^+(v)~.\nn
\eea

In the bridge  representation, the system of self-duality equations \p{E7}
is equivalent to the single equation
\be
\hcD^+_\da \left( e^v\Dp e^{-v}\right)\equiv\hcD^+_\da \Vp(v)=0~.
\ee

Let us introduce the nilpotent gauge condition for the bridge matrix
\be
v=\hT^{\da-}b^+_\da+(\hTm)^2b^\pp~,\q v^2=-(\hTm)^2b^{\da+}b^+_\da~,
\lb{sdgauge}
\ee
where $b^+_\da$ and $b^\pp$ are G-analytic matrix coefficients.
Note that the analogous nilpotent bridge representation has been used
recently in the $N=3$ SYM-theory \cite{NZ}.

Let us compose  harmonic connection $\Vp(v)$
\bea
&&e^v\Dp e^{-v}=-\hTpa b^+_\da+\hTma\left(-\Dp b^+_\da-\hT^+_\da b^\pp+
{1\over2}\hTpb\{b^+_\da,b^+_\db\}\right)\\
&&+(\hTm)^2\left(\Dp b^\pp+{1\over2}\{b^{\da+},\Dp b^+_\da\}+
\hTpa[b^+_\da,b^\pp]+{1\over2}\hTpa\{b^+_\da,b^{\db+}b^+_\db\}\right),\nn
\eea
where we have used the identity
\be
\{b^+_\da,b^{\db+}b^+_\db\}={1\over3}[b^{\db+},\{b^+_\da,b^+_\db\}]~.
\ee

The dynamical analyticity equation
\be
 \Vp(v)=-\hTpa b^+_\da
\ee
yields the following analytic relations:
\bea
&&\Dp b^+_\da+\hT^+_\da b^\pp-
{1\over2}\hTpb\{b^+_\da,b^+_\db\}=0~,\lb{hsd2}\\
&&\Dp b^\pp+{1\over2}\{b^{\da+},\Dp b^+_\da\}+
\hTpa[b^+_\da,b^\pp]+{1\over6}\hTpa[b^{\db+},\{b^+_\da,b^+_\db\}]=0~.
\nn
\eea

 Let us use the intermediate decomposition
of (2,2)-analytic superfields in $\hTpa$
\bea
&&b^+_\da=\beta^+_\da+\hTpb C_{\da\db}+(\hTp)^2\eta^-_\da~,\\
&&b^\pp=B^\pp+\hTpa\gamma^+_\da+(\hTp)^2B~,
\eea
where all coefficients are polynomials in $\Tpa$. The (2,2)-equations
for $b^+_\da$ and $b^\pp$ yield (2,0)-equations for these coefficients,
for instance,
\bea
&&\dpp \beta^+_\da=0~,\q\dpp B^\pp=0~,\lb{hom}\\
&&\dpp C_{\da\db}+\Tpb\pbdb\beta^+_\da+\varepsilon_{\da\db}B^\pp
-{1\over2}\{\beta^+_\da,\beta^+_\db\}
=0~.
\eea

The linear harmonic equations for $\beta^+_\da$ and $B^\pp$ can be solved
explicitly. The inhomogeneous linear harmonic equation for $C_{\da\db}$
contains the composed source calculated at the previous stage, so it is
also solvable, in principle. The inhomogeneous harmonic equations for
other (2,0)-coefficients $\eta^-_\da, \gamma^+_\da$ and $B$ can be derived
and solved analogously. It should be stressed that self-dual equations
\p{E7} and \p{hsd2} are convenient for the search of dimensionally reduced
solutions.

\setcounter{equation}0
\section{\lb{E}Harmonic constructions of self-dual solutions}

The alternative harmonic formalism for the  self-dual SYM-solutions has
been considered in Ref.\cite{DO}. This formalism harmonizes one of the
space  groups: $SU_\R(2)$ acting on the indices $\da$.

In order to compare alternative harmonic approaches to the $N=2$
self-dual equations we shall consider the non-covariant procedure
of identification $SU_R(2)=SU_A(2)$ for solutions of these equations.
Let us identify also the corresponding spinor indices $\da\equiv k,
\db\equiv l\ldots$ The $N=2$ superspace coordinates have the following
form in this notation:
\be
z^M=(y^{\alpha k}, \Theta^\alpha_k, \hT^k_l)~.
\ee

Using the single set of the $SU(2)/U(1)$ harmonics
\be
u^A_i\equiv \varepsilon^{AB}u_{Bi}~,\q A, B=\pm
\ee
one can consider the harmonic projections of central, analytic and chiral
coordinates, correspondingly:
\bea
&&y^{\alpha \A}=u^\A_ky^{\alpha k}~,\q Y^{\alpha \A}=y^{\alpha \A}-
\Theta^{+\alpha}\hT^{-\A}-\Theta^{-\alpha}\hT^{+\A}\\
&&y^{\alpha \A}_r=y^{\alpha \A}+
\Theta^{+\alpha}\hT^{-\A}-\Theta^{-\alpha}\hT^{+\A}\\
&&\Theta^\alpha_\A\equiv\varepsilon_{\A\B}\Theta^{\B\alpha}=u^k_\A
\Theta^\alpha_k~,\\
&&\hT^\B_\A\equiv\varepsilon_{\A\C}\hT^{\B\C}=u^l_\A u^\B_k\hT^k_l~,
\eea

The corresponding projections of the flat derivatives are
\bea
&&\partial_{\alpha\A}~,\q \cD_\alpha^\A~,\q\hcD^\A_\B~,\\
&&\{\cD_\alpha^\A,\hcD^\C_\B\}=-\varepsilon^{\A\C}\partial_{\alpha\B}.
\eea

It the central basis, one can obtain the following representations of the
$N=2$ covariant self-dual derivatives
\be
\nabla_\alpha^\pm=\cD^\pm_\alpha~,\q\hat\nabla_-^\pm=h^{-1}\hcD_-^\pm h~,
\q h^{-1}\partial_{\alpha-} h~,\lb{altrep}
\ee
where $h(z,u)$ is the chiral self-dual bridge
\be
\cD^\pm_\alpha h=0~.
\ee

The harmonic connection for the self-dual bridge is the basic
chiral-analytic prepotential of this formalism
\bea
&&h\cD^\pp h^{-1}=v^\pp~,\\
&&(\partial_{\alpha-}, \cD^\pm_\alpha, \hcD^\pm_-)v^\pp=0~.\lb{sdhs}
\eea
Stress that self-dual solutions possess the combined  analyticity. The
self-dual prepotential $v^\pp$ can be treated as unconstrained matrix
function of coordinates $y^{\alpha+}, u^\pm_i$ and $\hT^+_\pm$ which
parametrizes the general $N=2$ self-dual solution.

In the gauge group $SU(2)$, the simple solvable Ansatz for the self-dual
prepotential can be choosen
\be
(v^\pp)_i^k=u^{+k}u^+_i b^0 +(u^{+k}u^-_i+u^{-k}u^+_i)b^\pp~,
\ee
where $b^0$ and $b^\pp$ are real chiral-analytic functions. The bridge
matrix for this Ansatz can be calculated via harmonic quadratures
\cite{Zu}. In the simple example of this parametrization with $b^0=0$ and
\be
b^\pp=Y^{\alpha+}Y^{\beta+}\rho_\ab+\hT^+_+Y^{\alpha+}u^+_l
\lambda_\alpha^l
\ee
the solution depends on constant tensor and spinor coefficients.

It is not difficult to relate self-dual representations for different
methods of harmonization of superfield equations
\be
e^{-v}\hcD^+_-e^v=h^{-1}\hcD^+_-h~,\q \cD^-_\alpha(e^{-v}\hcD^+_-e^v)=
h^{-1}\partial_{\alpha-}h~.
\ee

\setcounter{equation}0
\section{\lb{G}Self-dual static $N=2$ solutions}

Consider now the subsidiary conditions in BPS-equations \p{C6} $r= p$ then
these equations can be transformed  to the following simple constraint:
\bea
&&-2i A_t+W+\bar{W}\equiv \left((\bDP\DP)+(\bDP)^2
-(\DP)^2\right)\Vm\nn\\&&
=-{1\over2}(D^{\alpha+}-\bar{D}^{\alpha+})(D_\alpha^+ -\bar{D}_\alpha^+)
\Vm=0~.\lb{bps1}
\eea
One can consider also the equivalent 2nd order constraints on $\Vm$ using
the transformation $\DPa\rightarrow e^{i\rho/2}\DPa$.

It is convenient to introduce the  new pseudoreal spinor coordinates
of the static harmonic superspace
\bea
&\Theta^{\alpha\pm}\equiv{1\over\sqrt{2}}(\theta^{\alpha\pm}+
\bar\theta^{\alpha\pm})~,\q\hT^{\alpha\pm}\equiv{1\over\sqrt{2}}
(\theta^{\alpha\pm}-\bar\theta^{\alpha\pm})~,&
\lb{rot3}\\
&(\Theta^{\alpha\pm})^\dagger =\Theta_\alpha^\pm~,\q
(\hT^{\alpha\pm})^\dagger =-\hT_\alpha^\pm~.&
\eea

The corresponding transformed  spinor derivatives
\bea
&&\cD^\pm_\alpha\equiv {1\over\sqrt{2}}(D^\pm_\alpha-\bar{D}^\pm_\alpha
)~,\q(\cD^\pm_\alpha)^\dagger=-\cD^{\alpha\pm}~,\lb{rot1}\\
&&\hcD^\pm_\alpha=-{1\over\sqrt{2}}(D^\pm_\alpha+\bar{D}^\pm_\alpha)~,\q
(\hcD^\pm_\alpha)^\dagger=\hcD^{\alpha\pm}~.
\eea
 have the following algebra:
\bea
&&\{\cD^+_\alpha,\hcD^-_\beta\}=-\{\hcD^+_\alpha,\cD^-_\beta\}=
-\partial_\ab~,\lb{stalg}\\
&&\{\cD^+_\alpha,\cD^-_\beta\}=
\{\hcD^+_\alpha,\hcD^-_\beta\}=0~.
\eea

Consider the $3D$-covariant representation of the Euclidean
superspace coordinates  \p{eucl1}
\be
y^\adb~\Rightarrow~y^\ab+\varepsilon^\ab y_4~.
\ee

Stress that the  algebra of spinor derivatives \p{stalg} arises also in
the dimensional reduction $y_4=0$ of the Euclidean $N=2$ harmonic
superspace (see Appendix). It is clear that the transformation \p{rot3}
connects the equivalent $3D$ subspaces of the Minkowski and Euclidean
types of harmonic superspaces. Thus, the (3,1)-equation \p{bps1} is
equivalent to the $3D$ limit of the Euclidean $N=2$ self-duality equation
\p{E7}
\be
(\cD^+)^2\Vm=0~.
\ee

The component static self-dual solutions con be obtained in the following
gauge:
\bea
&&\cV^\pp={1\over2}\hT^{\alpha+}\hT^+_\alpha c(x)+{i\over2}
\Theta^{\alpha+}\hT^+_\alpha a_t(x)+i\Theta^{\alpha+}\hT^{\beta+}a_\ab(x)
\nn\\
&&-\hT^{\alpha+}\hT^+_\alpha \Theta^{\beta+}u_k^-\Psi^k_\beta(x)~,
\eea
where all fields are Hermitian
\be
(c, a_t, a_\ab, \Psi^k_\alpha)^\dagger=(c, a_t, a^\ab, \Psi_k^\alpha)~.
\ee

One can derive the component $3D$ self-dual equations
\bea
&&F_\ab=-\nabla_\ab a_t~,\\
&&\nabla^\ab\nabla_\ab\,c=2[a_t,[a_t,c]]-\{\Psi^\alpha_k,\Psi^k_\alpha\}~,
\\
&&\nabla_\bg\Psi^{\gamma k}+{i\over2}[a_t,\Psi^k_\beta]=0~,
\eea
where
\be
\nabla_\ab\equiv\partial_\ab-i[a_\ab,]~,\q
F_\ab=\partial_\ar a^\rho_\beta+\partial_\br a^\rho_\alpha
-i[a_\ar,a_\beta^\rho].
\ee
Note that the static gauge field strength and field $a_t$ are connected
by the self-dual Bogomolnyi equation.

The superfield analysis of the static self-duality equations
can be made by analogy with the analysis of Euclidean $4D$ self-dual
equations in Sect. 4. The static limit of the bridge representation is
\be
A_\alpha^\pm=0~,\q\hat{A}_\alpha^+(v)=e^{-v}\hcD_\alpha^+e^v~,\lb{vrepst}
\ee
where $v$ is the static self-dual bridge $\cD^+_\alpha v=0$.

By analogy with the 4D self-duality equation one can use
the formal identification of all groups $SU(2)$ in the $3D$ Euclidean
harmonic superspace
\be
y^\ab\rightarrow y^{ik},\q\Theta^\alpha_l\rightarrow \Theta^k_l,\q
\hT^\alpha_l\rightarrow \hT^k_l~.
\ee

It is clear that one can use harmonic projections of the
superspace coordinates
\bea
&&y^{\A\B}_r=u^\A_ku^\B_ly^{kl}_r~,\\
&&\Theta^\B_\A=u^k_\A u_l^\B
\Theta^l_k~,\q\hT^\B_\A=u^l_\A u^\B_k\hT^k_l~.
\eea

The covariant derivative $\nabla^l_k$ are flat in the chiral
self-dual representation. The corresponding bridge representation
of the $3D$ self-dual covariant derivatives has the following form:
\be
h^{-1}\frac{\partial}{\partial y^{--}_r}h~,\qq
h^{-1}\frac{\partial}{\partial \hT^-_\A}h~.
\ee

The prepotential $v^\pp=h\Dp h^{-1}$ for these solutions depends on the
(1+2) analytic coordinates $y^\pp_r$ and $\hT^+_\A$ only. It is clear that
these solutions can be interpreted as the dimensionally reduced self-dual
$4D$ solutions \p{sdhs}.
\vspace{0.5cm}

{\bf Acknowledgement.}
Author is grateful to E. Ivanov and J. Niederle for discussions.
This work is supported in part by grants RFBR 99-02-18417,
RFBR-DFG-99-02-04022, INTAS-2000-254 and NATO PST.CLG 974874,
and by the  Votruba-Blokhintsev program in LTP JINR.

\appendix
 \def\theequation{A.\arabic{equation}}
\setcounter{equation}0
\section{\lb{F} Appendix}

{\bf Harmonic-superspace coordinates in $D=(3,1), N=2$ superspace}\\

The $SU(2)/U(1)$ harmonics  \cite{GIK1} parametrize the sphere $S^2$. They
form an $SU(2)$ matrix $u^\pm_i$ and are defined modulo $U(1)$.

The $SU(2)$-invariant harmonic derivatives act on the harmonics
\be
[\dpp ,\dm ]=\partial^0
~,\q [\partial^0 ,\partial^{\s\pm\pm}]=\pm 2\partial^{\s\pm\pm}~.
\ee

The special $SU(2)$-covariant conjugation of harmonics preserves the
$U(1)$-charges
\be
\widetilde{u^\pm_i}=u^{\pm i}\;,\q \widetilde{u^{\pm i}}=-u^\pm_i
~.\lb{conj1}
\ee
On the harmonic derivatives of an arbitrary harmonic function $f(u)$ this
conjugation acts as follows
\be
\widetilde{\partial^{\s\pm\pm} f}=\partial^{\s\pm\pm}\widetilde{f}
\;.\lb{conj2}
\ee

Let us consider the coordinates of the $N=2$ superspace $M(3,1|8)$
over the $D=(3,1)$ Minkowski space
\bea
&&z^M=(x^\adb~,\theta^\alpha_k~,\bar\theta^{\da k})~,\\
&&(x^\adb)^\dagger=x^\bda~,\q(\theta^\alpha_k)^\dagger=\bar\theta^{\da k}
~,
\eea
where $\alpha, \da$ are the $SL(2,C)$ indices.

One can define the Minkowski analytic harmonic superspace with 2 coset
harmonic dimensions $u^I_i$ and the following set of 4 even and (2+2) odd
coordinates:
\bea
&&\zeta =(x^\adb_\A , \tpa, \btpa)
\;,\nn\\
&&x^\adb_\A=x^\adb-i(u^-_ku^+_l+u^-_lu^+_k)\theta^{\alpha k}\bar
\theta^{\dot\beta l}\lb{F5}\\
&&
\theta^{\alpha\pm}=\theta^{\alpha k} u_k^\pm~,\q\bar\theta^{\dot\alpha\pm}
=\bar\theta^{\dot\alpha k}u_k^\pm~.\nn
\eea
This superspace is covariant with respect to the $N=2$ supersymmetry
transformations
\be
\delta x^\adb_\A=2i\tpa\bar\epsilon^{\dot\beta k}u_k^-
+2i\btpb\epsilon^{\alpha k}u_k^-~,\q \delta \theta^{\alpha\pm}=
\epsilon^{\alpha k}u_k^\pm~,\q\delta\bar\theta^{\dot\beta\pm}=
\bar\epsilon^{\dot\beta k}u_k^\pm~.
\ee

The conjugation of the odd analytic coordinates has the following
form:
\be
\theta^{\alpha\pm}~\rightarrow~\bar\theta^{\dot\alpha\pm}~,\q
\bar\theta^{\dot\alpha\pm}~\rightarrow~-\theta^{\alpha\pm}
\lb{conj3}
\ee
and coordinates $x^\adb_\A$  are real.

The corresponding CR-structure involves the derivatives
\be
\DPa,\;\bDPa,~\Dp
\lb{F7}
\ee
which have the following explicit form in these coordinates:
\bea
&& \DPa=\pPa~,\q\bDPa=\bpPa \;,\lb{F8}\\
&&D^{++}=\dpp
-i\tpa\btpb\padb+\tpa\pPa+\btpa\bpPa~,
\lb{F9}
\eea
where the partial derivatives satisfy the following relations:
\be
\pada x^\bdb=2\delta^\beta_\alpha\delta^\db_\da~,\q
\pPa\tmb=\delta^\beta_\alpha~,\q \bpPa\btmb=\delta^\db_\da~.
\ee

One can construct also all harmonic and Grassmann derivatives in these
coordinates
\bea
&&\Dm=\dm
-i\tma\btmb\padb+\tma\pMa+\btma\bpMa~,\\
&& \DMa =-\pMa +i\btmb\padb~,\q
\bDMa =-\bpMa -i\tma\pada~,\\
&&\pMa\tpb=\delta^\beta_\alpha~,\q \bpMa\btpb=\delta^\db_\da~.
\eea

It is useful to write down the properties of these derivatives
with respect to Hermitian conjugation
\bea
&&(\Dp, \Dm) \rightarrow -(\Dp, \Dm)~,\q D^\pm_\alpha \rightarrow
\bar{D}^\pm_\da~,\q\bar{D}^\pm_\da\rightarrow -D^\pm_\alpha ~.
\eea

In the text, we use the following conventions:
\bea
&&\varepsilon^{ik}\varepsilon_{kl}=\delta^i_l~,\q
\varepsilon^{\alpha\gamma}\varepsilon_{\gamma\beta}=\delta^\alpha_\beta~,
\\
&&(\theta^\pm)^2={1\over2}\theta^{\alpha\pm}\theta^\pm_\alpha~,\q
(\bar\theta^\pm)^2={1\over2}\bar\theta_\da^\pm\bar\theta^{\da\pm}~,\\
&&(D^\pm)^2={1\over2}D^{\alpha\pm}D^\pm_\alpha~,\q
(\bar{D}^\pm)^2={1\over2}\bar{D}_\da^\pm\bar{D}^{\da\pm}~.
\eea
\vspace{0.5cm}

{\bf Harmonic-superspace coordinates in Euclidean superspace}\\

The Euclidean $N=2$ superspace $E(4|8)$ has the coordinates
\be
z^M=(y^\adb,\Theta^\alpha_k,\hat{\Theta}^{\da k}) \lb{eucor}
\ee
 with the alternative conjugation rules
\be
(y^\ada)^\dagger=y_\ada~,\q(\Theta^\alpha_k)^\dagger=\Theta^k_\alpha~,\q
(\hat{\Theta}^{\da k})^\dagger=\hat{\Theta}_{\da k}~,\lb{eucl1}
\ee
where $\alpha, \da$ and $k$ are the indices of the direct product of
three $SU(2)$ groups.

The Euclidean $N=2$ analytic-superspace coordinates can be defined as
follows
\bea
&&\zeta_E =(y^\adb_\A , \Tpa, \hTpa)
\;,\nn\\
&&y^\adb_\A=y^\adb-(u^{k-}u^+_l+u_l^-u^{k+})\Theta^\alpha_k
\hat\Theta^{l\db}\\
&&
\Theta^{\alpha\pm}=\Theta^\alpha_k u^{k\pm}~,\q\hat\Theta^{\dot\alpha\pm}
=\hat\Theta^{k\da}u^\pm_k~.\nn
\eea

The Euclidean spinor and harmonic derivatives have the following form in
these coordinates:
\bea
&&\cDpa=\partial/\partial \Tma\equiv d^+_\alpha~,\q\hcDpa=
\partial/\partial \hTma\equiv \hat{d}^+_\da~,\\
&&\cD^{\pm\pm}=\partial^{\pm\pm}
-\Theta^{\alpha\pm}\hat\Theta^{\db\pm}\padb+\Theta^{\alpha\pm}
d^\pm_\alpha+\hat\Theta^{\da\pm}\hat{d}^\pm_\da~,\\
&& \cDma =-d^-_\alpha +\hat\Theta^{\db-}\padb~,\q
\hcD^-_\da =-\hat{d}^-_\da -\Theta^{\alpha-}\pada~.
\eea

We shall use the following rules of conjugation for
 the Euclidean harmonized odd coordinates and derivatives:
\bea
&&(\Theta^{\alpha\pm})^\dagger=\Theta_\alpha^\pm~,\q
(\cD^\pm_\alpha)^\dagger=-\cD^{\alpha\pm}~,\\
&&(\hat\Theta^{\dot\alpha\pm})^\dagger=-\hat\Theta_\da^\pm~,\q
(\hcD^\pm_\da)^\dagger=\hcD^{\pm\da}~.
\eea

Our conventions for the bilinear combinations of the Euclidean spinors
are
\bea
&&(\Theta^\pm)^2={1\over2}\Theta^{\alpha\pm}\Theta^\pm_\alpha~,\q
(\hT^\pm)^2={1\over2}\hT^{\da\pm}\hT^\pm_\da~,\nn\\
&&
(\cD^\pm)^2={1\over2}\cD^{\alpha\pm}\cD^\pm_\alpha~,\q
(\hcD^\pm)^2={1\over2}\hcD^{\da\pm}\hcD^\pm_\da~.
\eea

\end{document}